\documentclass[prb,aps,twocolumn,superscriptaddress,showpacs]{revtex4}
\usepackage{graphicx}
\usepackage{amsmath}
\usepackage{amssymb}
\begin{document}
\title{Kinetic Monte Carlo simulations of oscillatory shape 
evolution for electromigration-driven islands}
\author{Marko Rusanen}
\affiliation{Laboratory of Physics, Helsinki University of Technology,
P.O.\ Box 1100, FI--02015 TKK, Espoo,
Finland}
\author{Philipp Kuhn}
\affiliation{Institut f\"ur Theoretische Physik,
Universit\"at zu K\"oln,
Z\"ulpicher Strasse 77,
D--50937 K\"oln,
Germany}
\author{Joachim Krug}
\email[]{krug@thp.uni-koeln.de}
\affiliation{Institut f\"ur Theoretische Physik,
Universit\"at zu K\"oln,
Z\"ulpicher Strasse 77,
D--50937 K\"oln,
Germany}

\date{\today}
\begin{abstract}
The shape evolution of two-dimensional islands under electromigration-driven
periphery diffusion is studied by kinetic Monte Carlo (KMC) 
simulations and continuum theory. The energetics of the KMC model is
adapted to the Cu(100) surface, and the continuum model is matched to the
KMC model by a suitably parametrized choice of the orientation-dependent
step stiffness and step atom mobility. At 700 K shape oscillations predicted
by continuum theory are quantitatively verified by the KMC simulations, 
while at 500 K qualitative differences between the two modeling approaches
are found.

\end{abstract}

\pacs{68.65.-k, 66.30.Qa, 05.45.-a, 05.10.Ln}

\maketitle

\section{Introduction}

The nonequilibrium evolution of two-dimensional nanoscale structures is a central 
theme in current surface science, which has been the focus of many
experimental and theoretical studies 
\cite{Jeong99,Giesen01,Michely04,PierreLouis05,Kodambaka06}.
On the theoretical side, a key challenge is to bridge the length and time
scales between individual atomic processes and the mesoscopic step morphologies
that they produce \cite{PierreLouis05,Voigt05}. A highly successful approach
to this problem treats the crystal steps as continuous entities, whose dynamics
is governed by the interplay between the adatom population on the terraces, 
and the processes at the steps which define effective boundary
conditions on the adatom density field \cite{Jeong99,PierreLouis05,Krug05}. 
This approach is well established when the steps are rough and sufficiently
close to equilibrium for thermodynamic considerations to apply at least
locally. However, atomistic details not captured in the continuum theory
must generally be expected to become important at low temperatures and
under strong driving \cite{Combe00,PierreLouis01,Liu02,Iguain03}.

In this paper we compare atomistic and continuous modeling strategies for the 
specific case of island electromigration \cite{Metois,Mehl00,PierreLouis00}.
The present interest in electromigration-induced surface morphology evolution
is motivated both by its importance for the reliability of integrated
circuits \cite{Suo03}, and by its potential for a rich variety of nonlinear
dynamic phenomena. Specifically, it has recently been shown within a continuum
approach that the motion and shape evolution of electromigration-driven
islands in the regime dominated by periphery diffusion (PD) can be surprisingly
complex when the crystal anisotropy of the mobility of atoms along the island
edge is taken into account \cite{Kuhn05a,Kuhn05b}. Depending on the island
size and the strength of the anisotropy, a range of different dynamic phases
involving spontaneous symmetry breaking as well as oscillatory and chaotic
shape evolution has been predicted. 

With regard to a possible experimental
observation of these phases, it is important to understand to what extent the
features of the continuum model survive the discreteness and stochasticity
introduced into the dynamics on the atomic level. 
We have therefore conducted extensive Kinetic Monte Carlo (KMC) simulations 
of island electromigration, using a realistic model of the Cu(100)
surface \cite{Heinonen99}. At high temperatures ($T = 700$ K) we find good
agreement between the continuum and discrete models, and report for the first
time (to the best of our knowledge) a case of oscillatory shape evolution
within a fully stochastic KMC simulation. At lower temperatures the
oscillations persist, but additional features not covered by the continuum
description appear. The discrete and continuum models used in this work are 
described in the next Section. Results at high and low temperature are 
presented in Sect.\ref{Results}, and a concluding discussion can be found
in Sect.\ref{Conclusions}.

\section{Models}

\subsection{Kinetic Monte Carlo model}

Our KMC simulations are based on a lattice-gas model \cite{Heinonen99} 
with energetics obtained from effective medium theory
\cite{Merikoski97} (EMT). The energy barriers of the model
and their relative ordering are in good agreement with experimental 
data for the Cu(001) surface \cite{Giesen01}. It should be noted that,
within the EMT description, 
barriers on the Ag(001) and Ni(001) surfaces are very similar to those 
of Cu(001) up to an overall scaling factor \cite{Merikoski97}.
The simulations were implemented using the rejection-free Monte Carlo
algorithm by Bortz {\it et al.} \cite{Bortz75} with a binary tree
structure \cite{Blue95}. This allowed us to reach time scales on the order
of seconds in physical time and island sizes up to 10 000 atoms within
reasonable computation time.
Our model is very similar to that used by Mehl \textit{et al.}\cite{Mehl00}, who also considered
the Cu(100) surface. However, working at somewhat lower temperatures and with smaller
islands, they did not observe the complex phenomena described in the present paper.

In principle, the KMC model includes all relevant one-particle processes and
their corresponding energy barriers as given in Ref.19.
However, since our goal is to compare the KMC simulations with the
continuum theory, we need to exclude some processes from the KMC
simulations. Since in the continuum model the total mass of an island is
conserved, we explicitly forbid all adatom detachment events from an island.
This does however not exclude a jump into a site without any nearest
neighbors (NN) but with at least one next nearest neighbor atom. This local
restriction permits both atoms to jump around corner sites and excitation of
an atom at the boundary with three NN's into an edge atom with one NN through
two jumps.
We also exlude vacancy diffusion inside the island, as this is not included
in the continuum model. 

It must be emphasized that these restrictions do not
violate the detailed balance condition in our simulations.
We have checked that the oscillatory shape evolution phenomena described
in this paper persist even when they are relaxed. Needless
to say, however,
the definition of the island size becomes problematic for long
times when detachment of atoms from the island is allowed for.

The KMC simulations are conducted as follows. Initially an island with a given
number of atoms $S$ is placed on a lattice in a square or rectangular
configuration.
The hopping rate $\nu$ of an atom to a vacant NN site on
the island edge is given by
\begin{equation}
\nu =\nu _{0}\exp \left[ -(E_{S} - \min (0,\Delta _{NN})E_{B})/
k_{\mathrm{B}}T \right],
\end{equation}
where the attempt frequency $\nu_{0} = 3.0\times 10^{12}$ s$^{-1}$ and the 
barrier for the jump of an atom at a straight edge is $E_{S} = 0.26$ eV. The
change in bond number $-3 \leq \Delta_{NN}\leq 3$ is the difference in the
number of NN bonds between the initial and final states with bond energy
$E_{B}=0.26$ eV. Thus, the model gives an {\it additional} kink rounding
barrier $E_{\mathrm{kr}}=0.26$ eV (compared to the diffusion along the
close-packed $\left[110\right]$ edges) and a total detachment barrier
$E_{\mathrm{det}}=0.52$ eV from a kink site onto a edge site.

The electromigration force is introduced by decreasing (increasing) the barrier for
hops along (against) the direction of the force 
by an amount $E_{\mathrm{bias}}$. In most of the simulations described here
the force was directed along the $x$-direction of the lattice.
Experimentally realizable values of the bias\cite{Mehl00,Kuhn05a} 
are on the order of $E_{\mathrm{bias}} \approx 10^{-5}$ eV 
for Cu(100) at a current density of $10^7$ A/cm$^{2}$. The values used in 
the simulations
are typically two orders of magnitude larger. However, at least in certain
regimes it is easy to extrapolate to realistic bias strengths by considering
larger islands (see the discussion below in Sect.\ref{Results}).

\subsection{Continuum model}

The continuum description of two-dimensional shape
evolution by PD is based on the continuity equation 
\cite{PierreLouis00,Kuhn05a,Kuhn05b}
\begin{equation}
\label{cont}
v_n + \frac{\partial}{\partial s} a^2 \sigma \left[
- \frac{\partial}{\partial s} (a^2 \tilde \gamma \kappa) + F_t 
\right] = 0
\end{equation}
for the normal velocity $v_n$ of the island boundary. Here $s$ denotes the
arclength, $a$ the distance between NN sites, $\sigma$ the step atom
mobility, $\tilde \gamma$ the edge stiffness, $\kappa$ the local curvature, 
and $F_t$ the tangential component of the electromigration force acting on
an edge atom. We take the force to be of constant strength $F_0$ and aligned
along the $x$-axis. With $\theta$ denoting the angle between the normal of
the island edge and the $y$-axis (counted positive in the clockwise
direction), this implies $F_t = F_0 \cos(\theta)$.

The crystal anisotropy of the surface enters the continuum model through the
dependence of $\tilde \gamma$ and $\sigma$ on the edge orientation $\theta$.
In previous work based on (\ref{cont}) 
the stiffness was assumed to be isotropic, and a simple model expression with
sixfold symmetry was used for the mobility \cite{Kuhn05a,Kuhn05b}.
Here we are more ambitious, and aim to adapt the functions $\tilde \gamma(\theta)$ 
and $\sigma(\theta)$ as closely as possible to the thermodynamics and kinetics
of the KMC model.
  
The orientation dependence of the step stiffness on fcc(100) surfaces,
in particular on Cu(100), has been of considerable recent
interest \cite{Dieluweit03,Stasevich04}. For our purposes it is sufficient
to use the simple expression
\begin{equation}
\label{stiff}
\tilde \gamma(\theta) = \frac{k_\mathrm{B} T}{a} \frac{(1 + m^2)^{3/2}}{m^2 + 
\sqrt{m^2 + (a/l_k)^2}}, \;\;\; m = \tan(\theta),
\end{equation}
where $l_k = (a/2) e^{\epsilon/k_{\mathrm{T}}}$ is the distance
between thermally excited kinks, and $\epsilon = E_B/2=0.13$ eV is the kink energy. 
Equation (\ref{stiff}) interpolates in a natural way between the
leading order terms obtained in a low temperature expansion
of the full expression for the two-dimensional Ising lattice gas  
for $m > 0$ and $m = 0$, respectively. For the close-packed orientation
$m = 0$ it reduces to the well-known relation \cite{Jeong99,Giesen01,Michely04} 
\begin{equation}
\label{stiff0}
\tilde \gamma (0) = \frac{k_\mathrm{B} T}{a^2} l_k = \frac{k_\mathrm{B} T}{2 a} 
e^{\epsilon/k_{\mathrm{T}}}
\end{equation}
while for $m \neq 0$ and $T \to 0$ one obtains the purely entropic low-temperature
expression \cite{Dieluweit03}
\begin{equation}
\label{entropy}
\tilde \gamma (\theta) = \frac{k_\mathrm{B} T}{a} \frac{(1 + m^2)^{3/2}}{m^2 + 
\vert m \vert}.
\end{equation} 
The transition
between the two regimes occurs, and the singularity of (\ref{entropy})
is cut off, when the concentration $m$ of forced kinks
becomes comparable to the concentration $a/l_k$ of thermal kinks. 

The anisotropy of the mobility is less well understood, and experimental
data are so far scarce \cite{Giesen04,Tao06}. It is known that 
$\sigma$ is isotropic in the absence of an additional kink rounding barrier 
\cite{Krug95,Liu02}. Here we extend a phenomenological expression previously
derived for the close-packed step orientation \cite{PierreLouis01,Kallunki03} 
to general orientations by writing
\begin{equation}
\label{mob}
\sigma(\theta) = \frac{a \nu_0 e^{-E_{\mathrm{det}}/k_{\mathrm B} T}}{(k_{\mathrm B} T)
[1 + (a/l_{\mathrm{eff}}(\theta)) p_{\mathrm{kr}}^{-1}]},
\end{equation}
where $p_{\mathrm{kr}} = e^{-E_{\mathrm{kr}}/k_{\mathrm{B}}T}$ is the kink
rounding probability. The \textit{effective kink spacing} $l_{\mathrm{eff}}$,
which replaces the thermal kink spacing $l_{k}$ at nonzero tilts, is defined
in terms of the stiffness (given by (\ref{stiff})) through the relation
\begin{equation}
\label{leff}
l_{\mathrm{eff}}(\theta) = (a^2/k_{\mathrm{B}} T) \tilde \gamma(\theta).
\end{equation}
This is motivated by the general expression for the roughness of a step configuration
$\zeta(x)$,
\begin{equation}
\label{rough}
\langle (\zeta(x) - \zeta(x'))^2 \rangle = \frac{k_\mathrm{B} T}{\tilde \gamma} \vert
x - x' \vert,
\end{equation}
which is valid for steps of arbitrary orientation \cite{Jeong99,Giesen01,Michely04}. 
On the basis of (\ref{rough}) we define the effective kink spacing as that
length scale at which the amplitude of step fluctuations becomes equal to the 
NN lattice spacing. Equating the right hand side of (\ref{rough}) to $a^2$ then
immediately leads to (\ref{leff}). 

The dependence of the mobility (\ref{mob}) on temperature
and orientation is fully specified by the values of the kink energy and the
kink rounding barrier, and reproduces correctly the behavior in various known limits.
For $\theta = 0$ the low temperature behavior of the mobility depends \cite{Kallunki03} 
on whether the kink rounding barrier is weak (in the sense of $E_\mathrm{kr} < \epsilon$)
or strong ($E_{\mathrm{kr}} > \epsilon$). In the first case $\sigma(0) 
\sim e^{-E_{\mathrm{det}}/k_\mathrm{B} T}$, while in the second case 
$\sigma(0) \sim e^{-(E_{\mathrm{det}} + E_{\mathrm{kr}} - \epsilon)/k_\mathrm{B} T}$.
For kinked steps $l_{\mathrm{eff}}/a$ is of order unity, and hence
$\sigma \sim e^{-(E_{\mathrm{det}} + E_{\mathrm{kr}})/k_\mathrm{B} T}$,
in accordance with Monte Carlo simulations \cite{Liu02}. For the case of a strong 
kink rounding barrier, which is realized in our system, the anisotropy of the mobility
[the ratio between the maximum and minimum values of $\sigma(\theta)$] is thus of the order
of $e^{\epsilon/k_\mathrm{B} T}$, which is comparable to the anisotropy of the stiffness.
A detailed comparison of (\ref{mob}) with numerical measurements of the mobility within the 
KMC model will be presented elsewhere.

The continuum model (\ref{cont}) can be made dimensionless by scaling lengths
with 
\begin{equation}
\label{lE}
l_E = \sqrt{a^2 \tilde \gamma(0)/F_0}
\end{equation} 
and time with 
\begin{equation}
\label{tE}
t_E = \frac{l_E^4}{[\sigma(0) \tilde \gamma(0) a^4]}.
\end{equation}
This implies in particular that increasing the electromigration
force is equivalent (up to a rescaling of time) to an increase
of the island area by the same factor. Generally speaking,
electromigration effects become important when the linear
island size is comparable to $l_E$. When presenting the results
for the continuum model the island size will be parametrized
by the dimensionless radius $\hat R$
of a circle of the same area, and time is measured in units
of $t_E$. For comparison with the KMC simulations, the dimensionless
radius of an islands of $S$ atoms is then defined by
$\hat R = \sqrt{S/\pi}/l_E$.

\begin{figure}
[ptb]
\begin{center}
\includegraphics[scale=0.55]%
{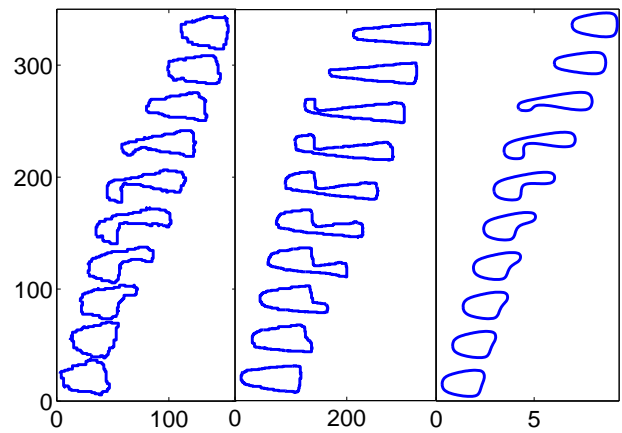}%
\caption{Time sequences of island configurations obtained from KMC simulations 
(left and middle columns) and numerical solution of the continuum equation
(rightmost column). Note that all configurations are displaced in the $y$-direction
linearly in time. Parameters are $T = 700$ K, $S = 1000$,
$E_{\mathrm{bias}} = 2 \times 10^{-3}$ eV (left column),  $T = 500$ K,
$S = 4000$, $E_{\mathrm{bias}} =  10^{-3}$ eV
(middle column), $T = 700$ K, $\hat R = 2.35$ (right column).}%
\label{Fig_configurations}%
\end{center}
\end{figure}

\section{Results}

\label{Results}

\subsection{700 K}

In Figure \ref{Fig_configurations} we show snapshots of the island shape
obtained from the KMC simulations (left panel at $T=700$ K, middle at
$T=500$ K) in comparison with
the continuum model results at $700$ K (right panel). 
Apart from the shape fluctuations associated with the discreteness of the KMC
model, the shape evolution at 700 K is seen to be very similar to that
predicted by the continuum model. The dimensionless radius corresponding to the
KMC simulation parameters is $\hat R = 1.56$, somewhat smaller than the value 
$\hat R = 2.35$ used in the continuum calculations.

In the KMC simulations an oscillation starts as random protrusions of the front edge of
the island boundary grow large enough to be amplified by the electromigration
force. The front edge effectively splits into two parts, and
a straight segment perpendicular to the field
appears on the island boundary, which is left behind as the protrusion
advances. In the reference frame of the island the standing segment moves backwards
and eventually reaches the back end of the island. At this point the island
has elongated to about twice its initial length. When the standing segment
disappears the island therefore contracts  
back into a quasi-stationary shape, from which the next oscillation is triggered.
The whole sequence of events is identical to that observed in the continuum model.
The formation of a standing boundary segment corresponds precisely
to the static facets found previously for islands of sixfold symmetry \cite{Kuhn05a}.  

It can be seen from Fig.\ref{Fig_configurations} that the shape oscillations break
the symmetry with respect to the horizontal (the direction of the electromigration force),
since the initial protrusion may appear either in the upper
or the lower part of the front edge. The continuum model is strictly deterministic.
Hence the direction of symmetry breaking is fixed by the initial condition, and each
oscillation event causes a vertical shift of the center of mass of the island in the 
same direction. This implies that the mean direction of 
island motion forms a nonzero angle with the direction of the force, and places this
mode of migration into the \textit{oblique oscillatory} (OO) phase that was previously identified
in the case of sixfold symmetry \cite{Kuhn05a}. In contrast, in the KMC 
simulations the direction changes randomly from one oscillation event to the next. At least
for the time sequences of limited length that were generated in our simulations, no 
correlations between subsequent events could be detected, and hence the island center of 
mass performs a random walk in the transverse direction.

Continuum calculations carried out over a range of dimensionless island radii show that the
OO phase is sandwiched between a narrow regime of oblique stationary (OS) 
motion (where islands of stationary shape 
move at a small angle with respect to the direction of the force)
for $2.16 < \hat R < 2.24$, and a regime where islands break up 
for $\hat R \geq 2.4$. For $\hat R \leq 2.16$ the motion is stationary and straight
in the direction of the force (the SS phase). Apart from a shift in the
locations of the phase boundaries, this is in agreement with 
the KMC simulations, where regimes of stationary and oscillatory behavior followed by 
island breakup were observed with increasing island size. The OS regime could not be 
unambiguously identified in the KMC simulations, possibly because the small obliqueness
in the direction of motion is easily lost in the fluctuations.

\begin{figure}
[ptb]
\begin{center}
\includegraphics[scale=0.6]%
{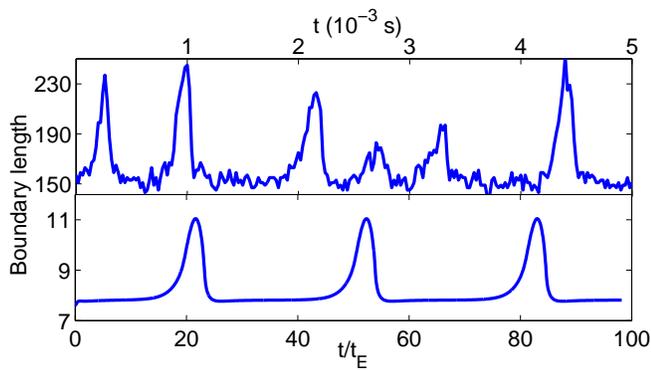}%
\caption{Oscillations of the boundary length obtained from KMC simulations
(upper panel) and the continuum model (lower panel). The conditions correspond
to the left and right columns of Fig.\ref{Fig_configurations}.}%
\label{Fig_oscillations}%
\end{center}
\end{figure}

Quantitatively the oscillation of the island shape can be monitored by various
observables such as the boundary length of the island, the average island velocity
or the center of mass coordinates. In Figure \ref{Fig_oscillations} we show the
evolution of the boundary length corresponding to the time sequences depicted in 
Figure \ref{Fig_configurations}. Despite a large amount of irregularity in the timing
of the oscillations observed in the KMC simulations, a typical period is clearly
visible, and the overall shape and amplitude of the individual events is quite
similar to those obtained from the continuum model. For the parameters of the 
KMC simulation the characteristic time scale is $t_E \approx 1.3 \times 10^{-4}$ s,
which is a factor 2.6 larger than the scale used to match the two time sequences
in Fig.\ref{Fig_oscillations}. Since $t_E \sim l_E^4$, this discrepancy is comparable
to that found above in the comparison of the effective island radius $\hat R$. In view
in particular of the uncertainties associated with the phenomenological expression    
(\ref{mob}) for the step atom mobility, the agreement between the two models must be 
considered remarkably good.

\begin{figure}
[ptb]
\begin{center}
\includegraphics[scale=0.4]%
{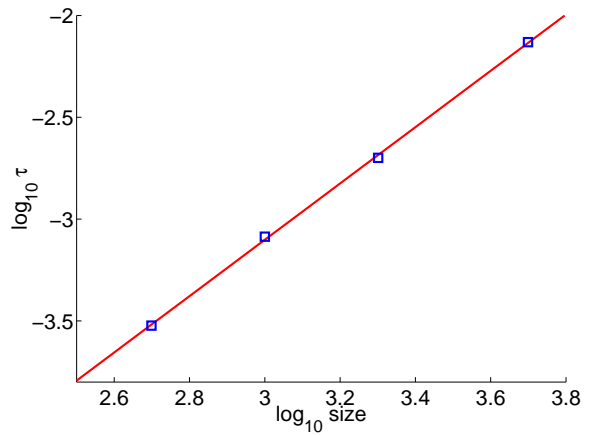}
\caption{Oscillation period (in units of seconds) 
measured in KMC simulations for islands of size $S = 500 - 5000$
and bias strengths $E_\mathrm{bias} = 4 \times 10^{-3}$ eV -- $4 \times 10^{-4}$ eV, such that
the dimensionless island radius is $\hat R = 1.56$ in all cases.}%
\label{Fig_tau}%
\end{center}
\end{figure}

Within the continuum model the behavior of the island depends only on the dimensionless
island size $\hat R$, defined relative to the characteristic length $l_E$ in (\ref{lE}). 
To check if this property carries over to the discrete model,
we performed a series of KMC simulations in which the island size and the strength of the 
bias were varied in such a way that the product $S E_\mathrm{bias}$ remained constant,
and which therefore correspond to a single value of the dimensionless island radius $\hat R$.  
As predicted by continuum theory, the same kind of oscillations depicted in the first column
of Fig.\ref{Fig_configurations} was observed for island sizes between $S = 500$ and $S = 5000$.
In addition, the oscillation period $\tau$ was found to scale with island size as a power law,
\begin{equation}
\label{Sz}
\tau \sim S^z
\end{equation} 
with $z \approx 1.4$ (Fig.\ref{Fig_tau}). The measured exponent $z$ is somewhat smaller
than the value $z = 2$ expected from continuum theory. There is however a substantial 
uncertainty in the estimates of $\tau$ both for large islands, where only a small number
of oscillations could be observed, and for small islands, where the oscillations are obscured
by noise.

\begin{figure}
[ptb]
\begin{center}
\includegraphics[scale=0.6]%
{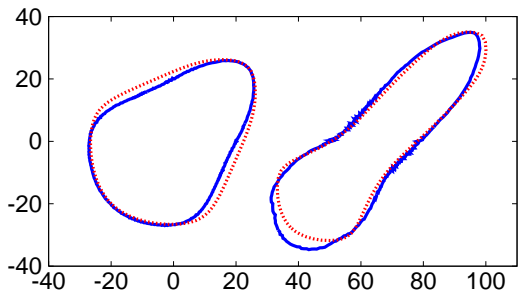}
\caption{Stationary island shapes obtained at $T = 700$ K with the electromigration force
directed along the lattice diagonal. Full blue lines show time-averaged KMC shapes
for island size $S = 2000$ and bias $E_\mathrm{bias} = 5 \times 10^{-4}$ eV (left) and
$E_\mathrm{bias} = 10^{-3}$ eV (right). Dotted red lines show stationary continuum shapes
of dimensionless radius $\hat R = 1.9$ (left) and $\hat R = 2.9$ (right).}%
\label{Fig_comp}%
\end{center}
\end{figure}

Exploratory simulations in which the direction of the electromigration force
was chosen to be different from the lattice axes indicate that the oscillatory behavior
generally persists. An exception is the case when the force points along the lattice diagonal,
where no oscillations are observed. Instead, the islands attain a characteristic shape which
becomes increasingly elongated with increasing island size or bias strength.
This behavior is well reproduced by the continuum model.  
In Fig.\ref{Fig_comp} the time-averaged shapes obtained from the KMC simulations are
compared to stationary shapes generated by the continuum model. The match between the
two sets of shapes is seen to be excellent, provided some adjustement of the effective
islands size is allowed for. The parameters of the KMC simulations in Fig.\ref{Fig_comp}
imply dimensionless radii of $\hat R = 1.106$ (left) and $\hat R = 1.56$ (right), 
which are again somewhat smaller than the values used for the continuum model.   

\subsection{500 K}

As shown in the middle column of Fig.\ref{Fig_configurations}, shape oscillations 
of a type similar to those seen at 700 K are observed also in KMC simulations at
500 K. Apart from the reduced amplitude of shape fluctuations, which is due to the larger
island size, the main difference between the left and middle columns is a pronounced
elongation of the island shape in all phases of the oscillation at 500 K.

\begin{figure}
[ptb]
\begin{center}
\includegraphics[scale=0.6]%
{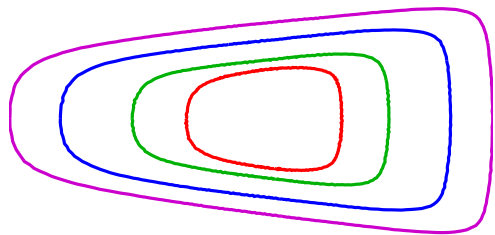}
\caption{Time-averaged stationary shapes for KMC islands of size $S = 500$, $1000$,
$2000$ and $3000$ atoms at $T = 500$ K and $E_{\mathrm{bias}} = 10^{-3}$ eV.
For this set of parameters oscillatory shape evolution sets in at $S = 4000$, see
the middle column of Fig.\ref{Fig_configurations}.}%
\label{Fig_shapes}%
\end{center}
\end{figure}

\begin{figure}
[ptb]
\begin{center}
\includegraphics[scale=0.6]%
{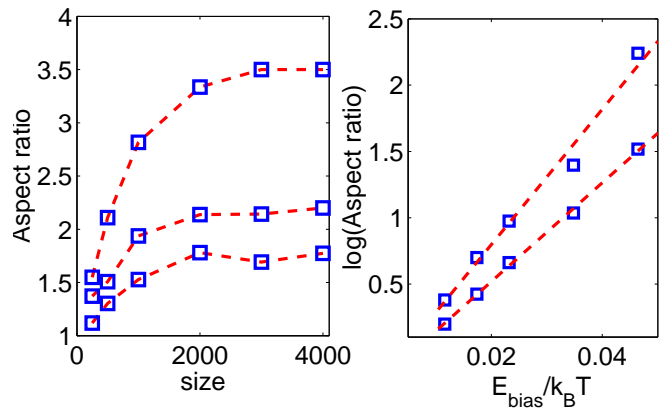}
\caption{Island aspect ratio, defined as the ratio of the maximal extensions
along and perpendicular to the field, for 
stationary islands obtained from KMC simulations at $500$ K. 
Left: the aspect ratio as a function of island size with
$E_{\mathrm{bias}} = 7.5 \times 10^{-4}$ eV, $1.0 \times 10^{-3}$ eV, and
$1.5 \times 10^{-3}$ eV (from bottom to top). The dashed lines are guides to
the eye. Right: the logarithm of the
aspect ratio as a function of bias for $S = 1000$ (lower line) and $S=4000$
(upper line). The dashed lines are linear fits to the data with the slopes
$37$ and $50$ for $S=1000$ and $S=4000$, respectively.}%
\label{Fig_aratio}%
\end{center}
\end{figure}

This elongation \cite{Mehl}
shows up also in the stationary shapes which are characteristic of smaller islands and
weaker bias (Fig.\ref{Fig_shapes}). A quantitative evaluation of the behavior of the
aspect ratio of stationary islands as a function of island size and bias strength is
shown in Fig.\ref{Fig_aratio}. The aspect ratio is seen to increase with increasing
island size up to some upper limit where it saturates. Increasing the island size
further eventually leads to the onset of oscillations, as shown
in the middle column of Fig.\ref{Fig_configurations}; at other parameter values
island breakup may also result. As was described above, during
an oscillation the front edge of the island splits up into two parts, one of which advances
while the other one stays behind. As a consequence, the island width is halved and the 
aspect ratio doubled at the point of maximal elongation. Between oscillations the
aspect ratio relaxes back to a value that coincides with the saturation value 
for large stationary islands shown in the left panel of Fig.\ref{Fig_aratio}.
In this sense the onset of oscillations can be seen as an attempt of the island
to increase its aspect ratio beyond the maximal value that can be realized by
stationary shapes.

The right panel of Fig.\ref{Fig_aratio} shows that the dependence of the island shape
on the strength of the electromigration bias is fundamentally different from the
dependence on island size: Rather than saturating, the aspect ratio of stationary islands
continues to grow with increasing bias, and no onset of oscillations is observed at least
for islands sizes $S < 4000$. The inequivalence between size and bias dependence implies
the breakdown of the continuum theory, which predicts that the islands shape should depend
only on the dimensionless parameter $S/l_E^2 \sim S E_\mathrm{bias}$. To give a quantitative
example, the data in Fig.\ref{Fig_aratio} show that an island of size $S = 2000$ driven by
a bias force of strength $E_\mathrm{bias} = 10^{-3}$ eV attains a stationary aspect ratio
$A \approx 2.2$, whereas for $S = 1000$ and $E_\mathrm{bias} = 2 \times 10^{-3}$ eV the
aspect ratio is twice as large. Within the continuum theory, the two parameter combinations
should lead to identical shapes. Indeed, continuum calculations carried out at $500$ K fail
to reproduce the elongated shapes seen in the KMC simulations. Instead, 
at the dimensionless radius $\hat R \approx 1.2$ corresponding to the cases described
above, one finds stationary shapes with an aspect ratio near unity. Larger islands undergo
OS motion for $2.1 < \hat R < 2.2$ and break up for $\hat R \geq 2.2$. 

Quantitatively the data for the aspect ratio $A$ in the right panel of Fig.\ref{Fig_aratio}
roughly follow an exponential relation 
\begin{equation}
\label{aspect}
A \sim e^{\alpha E_\mathrm{bias}/k_{\mathrm{B}} T}
\end{equation}
where $\alpha = 37$ and 50 for $S = 1000$ and 4000, respectively. While the origin of this
behavior so far eludes us, it is interesting to note that, in order of magnitude,
$\alpha \sim \sqrt{S}$. This indicates that the relevant energy scale governing the behavior
of the aspect ratio may be related to the energy gain associated with moving a row of atoms
(of length $\sim \sqrt{S}$) by one lattice spacing in the direction of the electromigration
force.

\section{Conclusions}

\label{Conclusions}

The most important outcome of our study is the observation of oscillatory island shape
evolution, a phenomenon first predicted on the basis of continuum theory \cite{Kuhn05a,Kuhn05b},
within a realistic KMC model of the Cu(100) surface. Since the KMC model has been previously
validated against experimental observations \cite{Giesen01}, we can now confidently pin down
the conditions under which electromigration-driven
shape oscillations should be realizable in the laboratory.
Assuming that a bias strength of the order of $E_\mathrm{bias} = 10^{-5}$ eV can be achieved
experimentally \cite{Mehl00}, extrapolation of the KMC results shown in Fig.\ref{Fig_tau}
indicate that oscillatory behavior of the kind shown in Fig.\ref{Fig_configurations} should
emerge at 700 K for islands containing about $2 \times 10^5$ atoms, corresponding to an 
effective radius of approximately 90 nm. The period of oscillations is expected to lie between
2 s and 40 s, depending on which value of $z$ is used in the relation (\ref{Sz}).   

It is instructive to compare the modes of island migration observed in the present study
of a surface with fourfold crystalline anisotropy to previous work on the continuum
model, where a sixfold anisotropy of the step atom mobility was assumed\cite{Kuhn05a}.
Generally speaking, we find that the parameter regime in which complex motion 
occurs (the OO phase discussed above in Sect.\ref{Results}) 
is less extended, and islands are more prone to breakup 
than in the sixfold symmetric case. Also, the spontaneous symmetry breaking 
associated with the OS phase (which could be detected here only in the deterministic
continuum dynamics) is more pronounced in the model with sixfold anisotropy, in the sense that the
angle between the direction of motion and the direction of the force is much larger\cite{Kuhn05a}. 
But it is reassuring
that the precise form of the anisotropies of stiffness and edge atom mobility 
seems to be inessential for the basic phenomenon of oscillatory shape evolution to emerge.

A detailed explanation of the breakdown of continuum scaling and the associated elongation
of island shapes observed in the simulations at 500 K is lacking at present, and 
we can offer only some rather general remarks. In previous studies
of island shape evolution deviations from the continuum approach were typically found when the
characteristic size of the islands becomes comparable to the spacing $l_k$ between thermal 
kinks \cite{Combe00,Liu02,Iguain03}. Since electromigration-dominated dynamics sets in at
island sizes comparable to $l_E$, it is instructive to consider the ratio
\begin{equation}
\label{ratio}
\frac{l_k}{l_E} = \sqrt{\frac{E_\mathrm{bias}}{2 k_\mathrm{B} T}} \; e^{\epsilon/2k_{\mathrm{B}} T},
\end{equation}
which is seen to increase with decreasing temperature and increasing bias strength. For the
standard value $E_\mathrm{bias} = 10^{-3}$ eV used in our work, the ratio (\ref{ratio})
increases from 0.27 at 700 K to 0.49 at 500 K, which makes it plausible that the continuum description
may become questionable in the latter case. According to this criterion,
for an experimentally realizable bias strength of $E_\mathrm{bias} = 10^{-5}$ eV the 
breakdown would be shifted to much lower temperatures, and the continuum description should
remain valid down to room temperature and below.

\section*{Acknowledgements}

This work was supported by DFG within project KR 1123/1-2, by the Academy of
Finland within its Center of Excellence program (COMP), and by the 
EU within project MagDot. We thank T. Ala-Nissila for fruitful discussions. 
JK is grateful to H. Mehl for useful
correspondence, and to the Laboratory of Physics at TKK
for its gracious hospitality during the early stages of this project.

\end{document}